\documentstyle[11pt,paspconf]{article}
\begin{document}

\title{IDENTIFICATION OF FAINT MARKARIAN GALAXIES WITH IRAS SOURCES}
\author{V. T. Ayvazyan}
\affil{\tt\footnotesize Special Astrophysical Observatory RAS Nizhnij Arkhys, Karachai-Cherkessia, 357147 Russia.
Electronic mail: ayvo@sao.ru\\
Armenian State Pedagogical Institute, Yerevan, Khanjian~5, 375010, Armenia\\}

\begin{abstract}

  The identification of IRAS sources with Faint Markarian galaxies
on the base of the Second Byurakan Survey were made. It is shown that
about 30\% of SBS galaxies are also IRAS sources.
The list of newly identified objects are presented.

\end{abstract}

\section{Introduction}

The Second Byurakan Survey (herafter SBS) was aimed to reach fainter limiting
magnitudes (Markarian \& Stepanian 1983,
1984a, 1984b; Markarian et al. 1985, 1986; Stepanian et al. 1988, 1990;
Stepanian 1994) in comparison with First Byurakan Survey (Markarian 1967).

Between 1974 and 1991 a total area of 1000 square degrees of the
sky were observed down to the limiting magnitude $19.\!\!^m\! 5$. This area is
confined by the strip defined by $7^{h} 40^{m} < \alpha < 17^{h} 15^{m}$,
$+49^\circ < \delta < +61^\circ$. A selection of more than 1700 galaxies
and about 1800 stellar objects with an excess ultraviolet emission
is the main result of the SBS survey (Stepanian 1994).

\section{Identification}

   Infrared data for nearly 500 SBS galaxies were obtained from the
NED\linebreak ({\small\it NASA/IPAC Extragalactic database}),
IPSS ({\small\it IRAS Point Source Survey}) and IFSS
({\small\it IRAS Faint Source Survey}).
About 50 SBS galaxies are newly identified with IRAS sources.
The difference of coordinates less than $1^\prime$ had been used
for preliminary identification. The presence of optical objects had been
analysed in the $1^\prime$ circle with the centres of IRAS coordinates.
Then the SBS galaxies as well as the other optical objects till the limit
of Digitized Palomar Sky Survey which were displased inside of mentioned
circle were analyzed.

The difference of coordinates for above mentioned 500 SBS IRAS
galaxies also may be used for preliminary identification. The plot of
this differences is shown in Fig.~1a. It is seen from Fig.~1a that nearly
80\% of objects are in the box of $20^{\prime\prime}$.

The similar plot for newly identified objects is shown in Fig.~1b.
The use of $20^{\prime\prime}$ criteria for newly identified objects shows that 20
objects may be identified as SBS IRAS sources. All this

galaxies are the only nearest objects around IRAS sources. There are 5
objects more, where the difference of coordinates is greater than $20^{\prime\prime}$,
but the optical SBS galaxy is the only nearest galaxy around IRAS source.
The remainder 25 SBS galaxies require additional identification.
Therefore, 25 SBS galaxies with the hight probability may be identified
as IRAS sources.

So, the total amount of SBS IRAS galaxies contains 525 objects,
that compose about 30\% of all SBS galaxies, 5\% of this sources are newly
identified. The list of these 25 objects is presented in the Table~1.

\acknowledgments

The autor thanks J. A. Stepanian for a formulation of the task and
repeated discussion of this paper and N. Serafimovich for her help
in draw up the article. This work were supported by the research
grants No. 97-02-17168 and 1.2.2.2 from the Russian Foundation for Basic
Research and from State Programm "Astronomy" respectively.

\begin{table}[h]
\begin{center}
\caption{The list of 25 newly identified objects.}
\vspace{1.3mm}
\begin{tabular}{lrlrr}
\hline
\multicolumn{1}{c}{SBS name} &
\multicolumn{1}{c}{IRAS name} &
\multicolumn{1}{c}{SBS name} &
\multicolumn{1}{c}{IRAS name}\\
\hline
0906+502   &  IRASF09067+5015  & 1339+559   &  IRAS\hspace{1.37mm} 13397+5555 \\
0933+524   &  IRASF09333+5227  & 1410+504   &  IRASF14107+5028 \\
0943+563A  &  IRASF09437+5620  & 1418+540   &  IRASF14187+5406 \\
1001+584   &  IRASF10016+5824  & 1423+600   &  IRASF14235+6000 \\
1020+610   &  IRAS\hspace{1.37mm} 10204+6100  & 1512+583   &  IRASF15122+5823 \\
1050+505   &  IRASF10502+5032  & 1519+508A  &  IRASF15195+5050 \\
1050+573   &  IRASF10507+5723  & 1528+577B  &  IRASF15288+5747 \\
1115+540A  &  IRASF11154+5401  & 1535+547   &  IRASF15353+5443 \\
1115+540B  &  IRAS\hspace{1.37mm} 11154+5401  & 1551+593B  &  IRASF15512+5923 \\
1115+588   &  IRASF11159+5853  & 1600+565   &  IRASF16005+5632 \\
1123+550   &  IRAS\hspace{1.37mm} 11236+5503  & 1609+490   &  IRASF16094+4902 \\
1125+581   &  IRAS\hspace{1.37mm} 11258+5806  & 1626+596   &  IRASF16263+5941 \\
1139+572   &  IRASF11392+5718  &\\
\hline
\end{tabular}
\end{center}
\end{table}

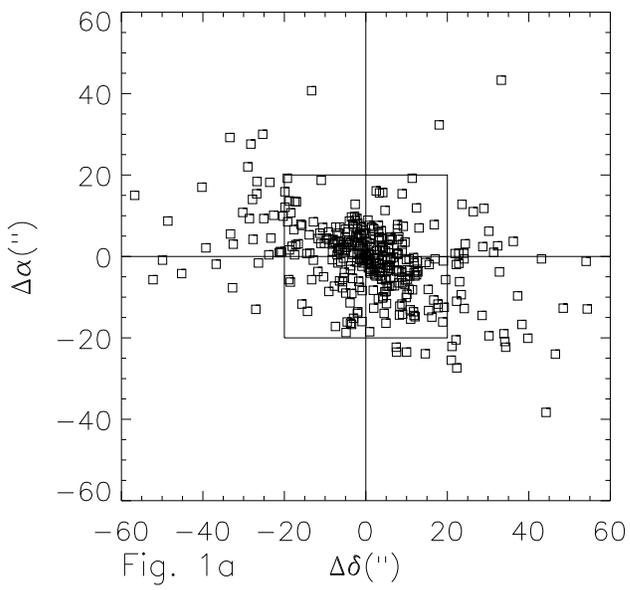
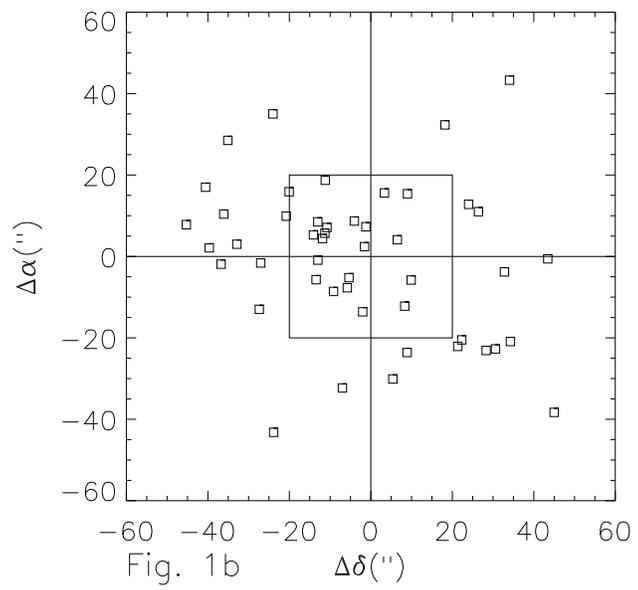
\end{document}